\documentclass{ws-mpla}
\usepackage[super]{cite}
\usepackage{graphicx}

\begin{document}

\markboth{Osetrin K.E., Makarenko A.N., Rybalov Yu.A.} {COSMOLOGICAL
MODELS WITH THE SPINOR AND NON-MINIMALLY INTERACTING SCALAR FIELD}

\catchline{}{}{}{}{}

\title{COSMOLOGICAL MODELS WITH THE SPINOR AND NON-MINIMALLY INTERACTING SCALAR FIELD}

\author{\footnotesize Osetrin K.E.\footnote{osetrin@tspu.edu.ru}}

\author{Makarenko A.N.\footnote{andre@tspu.edu.ru}}

\address{Tomsk State Pedagogical University,\\
Tomsk, Russian Federation}

\author{Rybalov Yu.A.\footnote{ribalovyua@tspu.edu.ru}}

\address{Tomsk State Pedagogical University,\\
Tomsk, Russian Federation,\\ National Research Tomsk State
University, \\ Tomsk, Russian Federation}

\maketitle

\pub{Received (Day Month Year)}{Revised (Day Month Year)}

\begin{abstract}

The solution to the current extending Universe problem, and the
description of all stages of evolution compels scientists to
consider various cosmological models. Scalar - tensor models are
rather simple and also allow us to  clearly define the separate
stages of evolution. Furthermore, other cosmological models are
reduced. Our work takes into consideration the non-minimally
interacted scalar field and the spinor field. The spinor field has
been considered to establish a better understanding of the stages of
evolution in our Universe.

\end{abstract}

\ccode{PACS Nos.: 04.50.Kd, 95.36.+x, 98.80.-k, 98.80.Cq}

\section{Introduction}

The problem of dark energy and dark matter is one of the main
challenges in modern cosmology. Astrophysical data indicates that
the observed universe is in an accelerated phase. This acceleration
could be caused by this so-called dark energy (see Refs.
~\refcite{lit1,lit2,lit3,lit4,lit5,lit6,lit7,lit8,lit9,lit10,lit11,Timoshkin}
for a recent review). These discoveries have set new serious
challenges for theoretical physics and have prompted speculation,
mostly based on phenomenological ideas which involve new dynamical
sources of gravity that act as dark energy, and/or various
modifications to general relativity. The spectrum of models, having
been postulated and explored in recent years, is extremely wide and
includes, in particular, Quintessence, K-essence, Ghost Condensates,
Dvali-Gabadadze-Porrati gravity, Galileon gravity, and $f(R)$
gravity Refs.  ~\refcite{lit12,lit13,lit14,Odintsov} (see Refs.
~\refcite{lit16} for detailed reviews of these and other models).
Such cosmological models tend to describe not only the accelerated
expansion at this stage of evolution, but also at all other stages
of evolution. All these models are divided into two main classes:
$f(R)$ theory and alike (see Refs.
~\refcite{lit12,lit16,lit17,lit17a,lit18}), and models which use
various objects: scalars, spinor, cosmological constant, liquid with
the difficult state equation.

The spinor fields have been used as a source of a gravitational
field in a number of works, and the interaction of the spinor and
scalar fields as a factor of inflation expansion has been
considered. Friedman-Robertson-Walker's (FRW) approach to the real
space-time is considered to be the closest, at the same time the
estimated change of a metrics conformal factor for the analysis of
behavior of the Universe in cosmological models is considered the
most often. In this work we will study cosmological models similar
to those offered in the works Refs.  ~\refcite{lit19,lit20}, but
with non-minimally interacted scalar field. Most of the
phenomenological models represent various modifications of
scalar-tensor theories. The models allowing non-minimal interaction
of scalar field derivatives and curvature are of particular
interest. As Amendola showed, the theory of such kind, cannot be
brought to a form of Einstein gravitation by conformal
transformation. Note that usually field equations in models with
non-minimal derivative interaction are the differential equations
above the second order. However, the order goes down to the second
in a special case when the kinetic term is connected only to
Einstein's tensor, i.e. $\kappa G_{\mu \nu } \phi ^{\mu } \phi ^{\nu
} $ (see, for example Ref.  ~\refcite{lit21}). In works Refs.
~\refcite{lit22,lit23} authors have studied cosmological scenarios
from non-minimal interaction of the derivative $\kappa G_{\mu \nu }
\phi ^{\mu } \phi ^{\nu } $ , concentrating on models with zero and
constant potential. According to the parameter choices, we have
obtained a variety of behaviors including the Big Bang, an expanding
universe without a beginning, a cosmological turnaround, an
eternally contracting universe, a Big Crunch, and a cosmological
bounce. In this cosmological model The non-minimal interaction of
gravitation and a matter (a scalar field) is considered in a
combination with the spinor field. In introduction the relevance of
the considered model is studied. In section 2 the task is formulated
and the main equations are given. In section 3 we will obtain the
FRW equations . In section 4 we will calculate solutions and conduct
asymptotic research solutions. In section 5, the results of our
research will be discussed.

\section{The field equations}

Let us consider a model with the action in the form:
\begin{equation} \label{EQ__1_}
S=\int  d^{4} x\sqrt{-g} \{ \frac{R}{8\pi } -(g_{\mu \nu } +\kappa
(R_{\mu \nu } -\frac{1}{2} Rg_{\mu \nu } ))\nabla ^{\mu } \phi
\nabla ^{\nu } \phi -V(\phi )-L_{D} \} ,
\end{equation}
 where $R$ is scalar curvature.

The Dirac Lagrangian $L_{D} $ of fermion mass field $m_{f} $ has the
form:
\begin{equation} \label{EQ__2_}
L_{D} =\frac{i}{2} \{ \bar{\psi }\Gamma ^{\mu } D_{\mu } \psi
-D_{\mu } \bar{\psi }\Gamma ^{\mu } \psi \} -m_{f} \langle \bar{\psi
}\psi \rangle -F(\psi \bar{\psi }).
\end{equation}
In the expression \eqref{EQ__2_}, $F(\psi \bar{\psi })$ describes
the potential of fermion field and $\bar{\psi }=\psi ^{\dag } \gamma
^{0} $ denotes the conjugate spinor. $\Gamma ^{\mu } =e_{a}^{\mu }
\gamma ^{a} $ are generalized Dirac-Pauly matrices in a curved
spacetime (where $e_{a}^{\mu } $ is tetrad).

A consideration of the spinor field is carried out in works Refs.
~\refcite{lit19,lit20}, we have used the necessary results. Let us
now consider a FRW universe with the flat spatial metric:
\begin{equation} \label{EQ__3_}
ds^{2} =-dt^{2} +a(t)^{2} (dx^{2} +dy^{2} +dz^{2} ).
\end{equation}
Einstein's equations can be written as
\begin{equation} \label{EQ__4_}
\frac{1}{8\pi } (R_{\mu \sigma } -\frac{1}{2} g_{\mu \nu } R)=T_{\mu
\nu } ,
\end{equation}
 where
$$T_{\mu \nu } =-(T_{f} )_{\mu \nu } +(T_{\phi } )_{\mu \nu } ,$$
($T_{f} )_{\mu \nu } $ is the energy-momentum tensor of the fermion
fields and $(T_{\phi } )_{\mu \nu } $ is the contribution of the
variation of the scalar field which interacts non-minimally. A
symmetric form of the energy-momentum of the fermion field is as
follows:
\begin{equation} \label{EQ__5_}
(T_{f} )_{0}^{0} =m_{f} \langle \bar{\psi }\psi \rangle +F(\psi
\bar{\psi }),
\end{equation}
\begin{equation} \label{EQ__6_}
(T_{f} )_{i}^{i} =F(\psi \bar{\psi })-\frac{\bar{\psi }}{2}
\frac{dF(\psi \bar{\psi })}{d\bar{\psi }} -\frac{dF(\psi \bar{\psi
})}{d\psi } \frac{\psi }{2} .
\end{equation}
The consequence of the equations of motion of the spinor fields is
given by:
\begin{equation} \label{EQ__7_}
\frac{d}{dt} \bar{\psi }\psi +3H\bar{\psi }\psi =0,\; \bar{\psi
}\psi =\frac{c}{a(t)^{3} } ,
\end{equation}
where $H=\frac{\dot{a}}{a} $. Self-interaction potential can be
written as $F=\sum _{n} \alpha _{n} (\bar{\psi }\psi )^{2n} $, where
$\alpha _{n} $ and n are constants.

A symmetric form of the scalar field energy-momentum tensor can be
obtained from \eqref{EQ__1_} in the form
\begin{eqnarray}
(T_{\phi } )_{\mu \nu } =\nabla _{\mu } \phi \nabla _{\nu } \phi
-\frac{1}{2} g_{\mu \nu } \nabla ^{\alpha } \phi \nabla _{\alpha }
\phi -g_{\mu \nu } V(\phi )+\kappa (\frac{1}{2} \nabla _{\mu }
\nabla _{\nu } (\nabla ^{\alpha } \phi \nabla _{\alpha } \phi )-\nonumber\\
-\frac{1}{2} R\nabla _{\mu } \phi \nabla _{\nu } \phi -\frac{1}{2}
(R_{\mu \nu } -\frac{1}{2} Rg_{\mu \nu } )\nabla ^{\alpha } \phi
\nabla _{\alpha } \phi -\frac{1}{2} g_{\mu \nu } {\rm \square
}(\nabla ^{\alpha } \phi \nabla _{\alpha } \phi )-\nonumber\\
-\frac{1}{2} g_{\mu \nu } R^{\alpha \beta } \nabla _{\alpha } \phi
\nabla _{\beta } \phi +2\nabla _{\alpha } \phi \nabla _{(\mu } \phi
R_{\nu )}^{\alpha } +\frac{1}{2} {\rm \square }(\nabla _{\mu } \phi
\nabla _{\nu } \phi )-\nonumber\\
-\nabla _{\alpha } \nabla _{(\mu } (\nabla _{\nu )} \phi \nabla
^{\alpha } \phi )+\frac{1}{2} g_{\mu \nu } \nabla _{\alpha } \nabla
_{\beta } (\nabla ^{\alpha } \phi \nabla ^{\beta } \phi )).
\label{EQ__8_}
\end{eqnarray}
We now write the equation of motion of the scalar field as:
\begin{equation} \label{EQ__9_}
{\rm \square }\phi +\kappa \nabla _{\mu } (\nabla _{\nu } \phi
(R^{\mu \nu } -\frac{1}{2} g^{\mu \nu } R))+V'(\phi )=0,\; V'(\phi
)=\frac{dV(\phi )}{d\phi } .
\end{equation}

\section{The field equations in the FRW}

We will choose the potential of a fermionny field $F(\bar{\psi }\psi
)$ in the following form:
\begin{equation} \label{EQ__10_}
F(\bar{\psi }\psi )=\alpha _{1} (\bar{\psi }\psi )^{2} +\alpha _{2}
(\bar{\psi }\psi )^{4} +\alpha _{3} (\bar{\psi }\psi )^{6} .
\end{equation}
 Einstein's equations in FRW will be as follows:
\begin{equation} \label{EQ__11_}
3\dot{a}^{2} =8\pi a^{2} (V(\phi )+\frac{m_{f} c}{a^{3} } +\alpha
_{1} \frac{c^{2} }{a^{6} } +\alpha _{2} \frac{c^{4} }{a^{12} }
+\alpha _{3} \frac{c^{6} }{a^{18} } )+4\pi (a^{2} -9\kappa
\dot{a}^{2} )\dot{\phi }^{2} ,
\end{equation}
\begin{eqnarray}
\dot{a}^{2} +2a\ddot{a}=-8\pi a^{2} (V(\phi )-\alpha
_{1} \frac{c^{2} }{a^{6} } -3\alpha _{2} \frac{c^{4} }{a^{12} }
-5\alpha _{3} \frac{c^{6} }{a^{18} } )-\nonumber\\
-4\pi \dot{\phi }^{2} (a^{2} +\kappa (\dot{a}^{2}
+2a\ddot{a}+4a\dot{a}(\dot{\phi })^{-1} \ddot{\phi })).
\label{EQ__12_}
\end{eqnarray}
The equation of motion of the scalar
field:
\begin{equation} \label{EQ__13_}
a^{3} \ddot{\phi }+3a^{2} \dot{a}\dot{\phi }-3\kappa (\dot{a}^{3}
\dot{\phi }+a\dot{a}(2\dot{\phi }\ddot{a}+\dot{a}\ddot{\phi
}))+V'(\phi )=0.
\end{equation}
 We will make replacements to simplify these equations:
\begin{equation} \label{EQ__14_}
\tau =ln\; a;\quad \quad \frac{d}{dt} =H\frac{d}{d\tau } .
\end{equation}
$$'=\frac{d}{d\tau } ,\; H=H(\tau ),\; \phi =\phi (\tau ).$$
\begin{equation} \label{EQ__15_}
X(\tau )=H^{2} ,\quad \quad Y(\tau )=\kappa X.
\end{equation}
 It is possible to choose two equivalent systems of the independent equations: \eqref{EQ__11_} and \eqref{EQ__12_}; \eqref{EQ__11_} and \eqref{EQ__13_},
 which will result in only one equation. We will consider the system of the equations \eqref{EQ__11_}, \eqref{EQ__13_}, we will receive the equation on $X(\tau )$:
\begin{eqnarray}
- \{ 8\pi \kappa (ce^{15\tau } m_{f} -c^{2} e^{12\tau } \alpha _{1}
-c^{4} e^{6\tau } \alpha _{2} -c^{6} \alpha _{3} -e^{18\tau }
V)(1+9X\kappa )+\nonumber\\
+e^{18\tau } (1-9X\kappa +54X^{2} \kappa ^{2} ) \} X'+8e^{18\tau }
\pi
V(2-24X\kappa +54X^{2} \kappa ^{2} )-\nonumber\\
-2\{(4ce^{15\tau } m_{f} \pi +3e^{18\tau } X+8c^{4} e^{6\tau } \pi
\alpha _{2} +16c^{6} \pi \alpha _{3} )\times \nonumber\\
(-1+3X\kappa )(-1+9X\kappa )+4e^{9\tau } \sqrt{\pi } X\kappa
\sqrt{-1+9X\kappa } \times \nonumber\\
 \sqrt{-3e^{18\tau } X-8\pi (ce^{15\tau } m_{f}
-c^{2} e^{12\tau } \alpha _{1} -c^{4} e^{6\tau } \alpha _{2} -c^{6}
\alpha _{3} -e^{18\tau } )V} V'\}=0. \label{EQ__16_}
\end{eqnarray}
We will notice that for $\alpha _{1} =\alpha _{2} =\alpha _{3}
=m_{f} =0$ the equation is similar to the equation formulated  in
work Ref.  ~\refcite{lit23}. In the case of a lack of potential of a
scalar field the equation \eqref{EQ__16_} shapes into a simple form:
\begin{eqnarray}
2(3e^{18\tau } Y+4e^{15\tau } \tilde{m}_{f} \pi +8e^{6\tau } \pi
\tilde{\alpha }_{2} +16\pi \tilde{\alpha }_{3} )(1-12Y+27Y^{2}
)+\nonumber\\
+(8\pi (1+9Y)(e^{15\tau } \tilde{m}_{f} -e^{12\tau } \tilde{\alpha
}_{1} -e^{6\tau } \tilde{\alpha }_{2} -\tilde{\alpha }_{3}
)+\nonumber\\
+e^{18\tau } (1-9Y+54Y^{2} ))Y'=0,\label{EQ__17_}
\end{eqnarray}
$$\tilde{m}_{f} =c\kappa m_{f} ,\; \tilde{\alpha }_{1} =c^{2} \kappa
\alpha _{1} ,\; \tilde{\alpha }_{2} =c^{4} \kappa \alpha _{2} ,\;
\tilde{\alpha }_{3} =c^{6} \kappa \alpha _{3} .$$ Thus the
independent system of the field equations in lack of potential of a
scalar field includes the equation \eqref{EQ__11_} and the equation
\eqref{EQ__17_}.

\section{Resolution of cosmological constant problem due to
coupled fermion}

The set of similar scalar models is considered in Ref.
~\refcite{lit16}, but in our work the model is not only considered
with the scalar, but also with the spinor field. We will contemplate
solutions of the field equations \eqref{EQ__11_}, \eqref{EQ__16_}.

Equation \eqref{EQ__11_} shows us that the existence of a special
value for H follows, namely at $H=\frac{1}{3\sqrt{\kappa } } $
coefficient at $\phi '^{2} $ addresses in zero. Furthermore,  it is
shown that, in this case, there is a solution.

\textbf{ I) Solution} $H={\kern 1pt} const{\kern 1pt} =\pm
\sqrt{1/(3\kappa )} $.

In this case the potential of a scalar field plays a role of a
cosmological constant $\Lambda $. $$V(\phi )={\kern 1pt} const{\kern
1pt} =\Lambda $$
\begin{equation} \label{EQ__18_}
a(t)=\beta e^{\frac{t}{\sqrt{3\kappa } } } ,\quad \quad \beta
-const,
\end{equation}
$$\dot{\phi }^{2} =\Lambda -\frac{1}{8\pi \kappa } +\frac{1}{\beta
^{18} } (c^{6} e^{-\frac{6\sqrt{3} t}{\sqrt{\kappa } } } \alpha _{3}
+c^{4} e^{-\frac{4\sqrt{3} t}{\sqrt{\kappa } } } \alpha _{2} \beta
^{6} +c^{2} e^{-\frac{2\sqrt{3} t}{\sqrt{\kappa } } } \alpha _{1}
\beta ^{12} -ce^{-\frac{\sqrt{3} t}{\sqrt{\kappa } } } m_{f} \beta
^{15} ),$$

\textbf{ II) Solution} $H={\kern 1pt} const{\kern 1pt} =\pm
\sqrt{1/(9\kappa )} $.
\begin{equation} \label{EQ__19_}
a(t)=\beta e^{\frac{t}{\sqrt{9\kappa } } } ,\quad \beta ,\gamma
-const,
\end{equation}
$$\dot{\phi }^{2} =\gamma e^{\frac{2t}{\sqrt{\kappa } } }
-\frac{1}{4\beta ^{18} } (9c^{6} e^{-\frac{6t}{\sqrt{\kappa } } }
\alpha _{3} +8c^{4} e^{-\frac{4t}{\sqrt{\kappa } } } \alpha _{2}
\beta ^{6} +6c^{2} e^{-\frac{2t}{\sqrt{\kappa } } } \alpha _{1}
\beta ^{12} -4ce^{-\frac{t}{\sqrt{\kappa } } } m_{f} \beta ^{15}
),$$
$$V(\phi )=\frac{1}{24\pi \kappa } -\frac{1}{\beta ^{18} }
(c^{6} e^{-\frac{6t}{\sqrt{\kappa } } } \alpha _{3} +c^{4}
e^{-\frac{4t}{\sqrt{\kappa } } } \alpha _{2} \beta ^{6} +c^{2}
e^{-\frac{2t}{\sqrt{\kappa } } } \alpha _{1} \beta ^{12}
-ce^{-\frac{t}{\sqrt{\kappa } } } m_{f} \beta ^{15} ).$$

We will now consider now solutions of the field equations in the
absence of scalar potential ($V(\phi )=0$), the equation
\eqref{EQ__11_} and \eqref{EQ__17_}. Asymptotic solutions are of
greater interest to us in the equation \eqref{EQ__17_} at small
values $a$ ($\tau \to -\infty $) and big $a$ ( $\tau \to +\infty $),
and as transition from $\tau \to -\infty $ to $\tau \to +\infty $.

\textbf{ A)} We will consider a case $\tau \to -\infty $, in this
limit the equation \eqref{EQ__17_} can be written down in the
following way ($Y=\kappa X=\kappa H^{2} $):
\begin{equation} \label{EQ__20_}
4(1-3Y)(1-9Y)-(1+9Y)Y'=0,
\end{equation}
For this equation there is a general solution:
\begin{equation} \label{EQ__21_}
Y_{1,2} =\frac{1}{6} (2+3\alpha e^{12\tau } \pm \sqrt{\alpha
e^{12\tau } (9\alpha e^{12\tau } +8)} ),\quad \alpha -{\kern 1pt}
const{\kern 1pt} .
\end{equation}
 The solution at $\tau \to -\infty $ takes the form
\begin{equation} \label{EQ__22_}
Y_{1,2} \approx \frac{1}{3} (1\pm \sqrt{2\alpha } {\kern 1pt}
e^{6\tau } )\to \frac{1}{3} .
\end{equation}
For the equation \eqref{EQ__20_} there are two partial solutions
$Y=1/9,\; Y=1/3$. The solution $Y=1/9$ -- stable, $Y=1/3$ --
unstable. There is a special value of $Y=-1/9$ for which $Y'$ does
not exist.
\begin{figure}[ph]
\begin{center}
\includegraphics[width=2.0in]{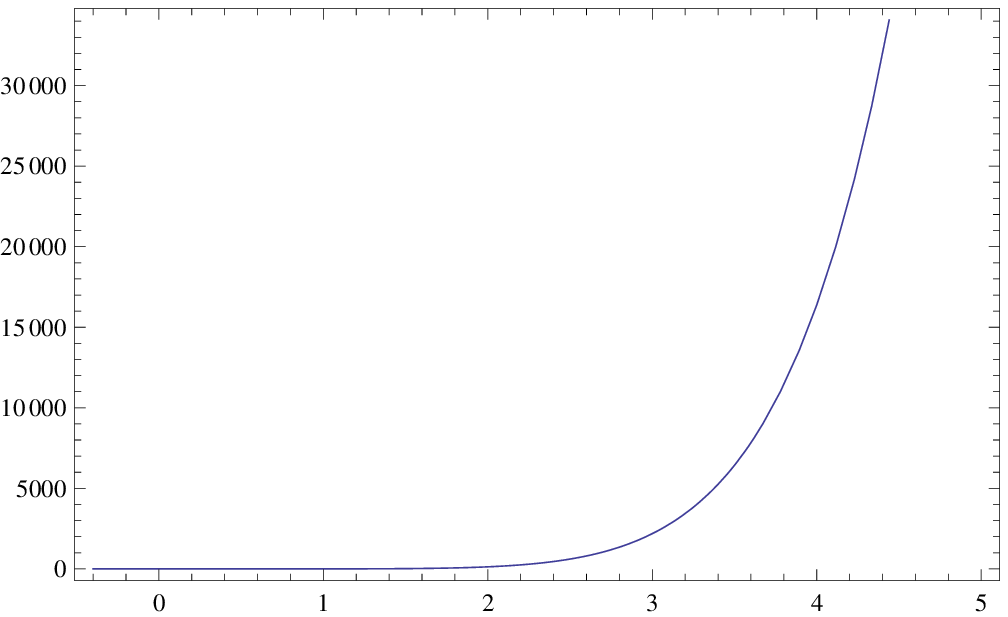}
\vspace*{8pt} \caption{Function $a'(a)$, $\tau \to -\infty$, $\kappa
>0$, $\alpha=1$ \protect\label{ris5}}
\end{center}
\end{figure}
At $\tau \to -\infty $ for $Y$ from an $(-\infty ,\; -1/9)$ interval
at approach for special value of $Y'$ tends to $-\infty $, for $Y$
from an $(0,\; 1/9)$ interval at approach for special value of $Y'$
tends to $+\infty $, for $Y$ from interval of $(1/9,\; +\infty )$
the solution tends to $Y=1/9$, $a(t)=\alpha
e^{\frac{t}{\sqrt{9\kappa } } } $. In the field of $Y\in (-\infty
,\; 0)$ the main contribution to the equation gives the summands of
the spinor field, generally summed with $\alpha _{3} $. If in the
equation \eqref{EQ__20_} we consider solutions of higher order of
smallness, then the most important are summands with $\alpha _{2}
,\; \alpha _{3} $ and the behavior of function $Y$ qualitatively
doesn't change. Thus, if we put the potential of the spinor field
equal to zero, we will receive a reversed diagram relative to the
$Ox$ axis. If we put $m_{f} ,\; \alpha _{2} ,\; \alpha _{3} =0$ we
will receive singular solutions $Y=0,\; Y=-1/9$. Note that the
results calculated in works Refs.  ~\refcite{lit22,lit23}, in our
case, are impossible, as in a limit at $\tau \to -\infty $ the
spinor field renders considerable influence. If we exclude the
spinor field, we will receive solutions given in the work above.

\begin{figure}[ph]
\centerline{\includegraphics[width=2.0in]{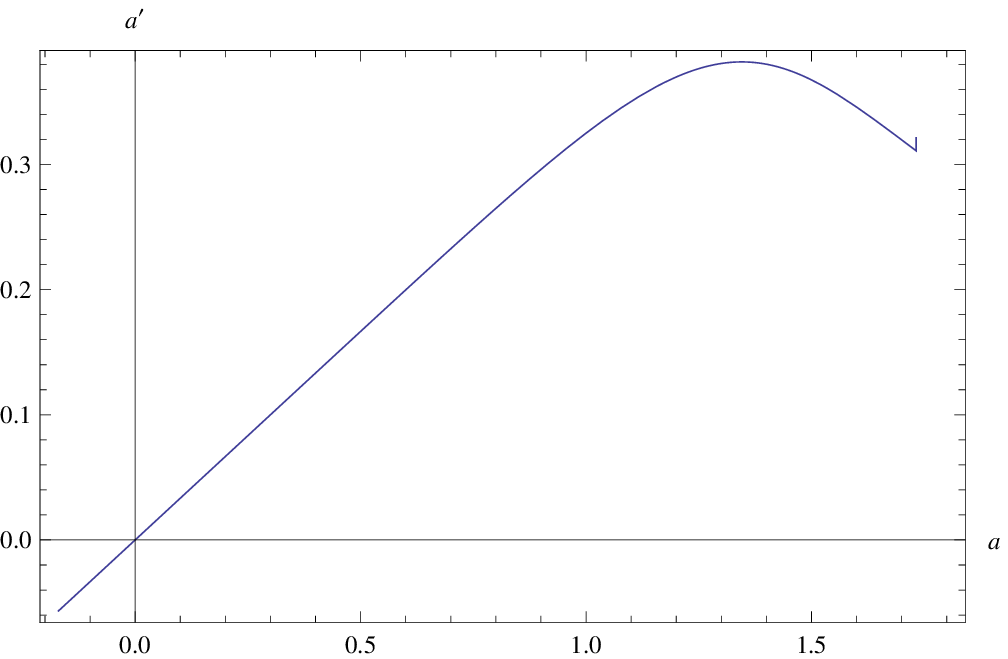}} \vspace*{8pt}
\caption{Function $a'(a)$, $\tau \to +\infty$, $\kappa >0$,
$\alpha=1$ \protect\label{ris6}}
\end{figure}

From the diagram (fig. \ref{ris5}) we can see that the accelerated
expansion is observed. Thus, while comparing diagram (fig.
\ref{ris5} and fig. \ref{ris6}) we can see the prevalence
(existence) of the summand spinor field, which causes faster
expansion.

\textbf{ B)} We will consider $\tau \to +\infty $ case. In this
limit the equation \eqref{EQ__17_} can be written down in the
following way:
\begin{equation} \label{EQ__23_}
6Y(1-3Y)(1-9Y)+(1-9Y+54Y^{2} )Y'=0,
\end{equation}
This equation is equivalent to the equation received in the work of
Ref.  ~\refcite{lit22} at $V(\phi )=0$, and, if we try to repeat the
calculation it is possible to receive similar solutions.

As it is possible to find the general solution of this equation.
\begin{equation} \label{EQ__24_}
Y=\frac{2}{9} +\frac{2^{1/3} e^{6\tau } }{9F(\tau )} -\frac{2^{1/3}
3\alpha }{F(\tau )} +\frac{2^{2/3} e^{-6\tau } F(\tau )}{18} ,
\end{equation}
\begin{equation} \label{EQ__25_}
F(\tau )=(-2e^{18\tau } -81\alpha e^{12\tau } +9\sqrt{\alpha
e^{18\tau } (8e^{12\tau } -27\alpha e^{6\tau } +972\alpha ^{2} )}
)^{1/3} ,
\end{equation}
where $\alpha $ - an integration constant. If $\tau \to +\infty $,
we have
$$F(\tau )\to -2^{1/3} e^{6\tau } (1-3\sqrt{2\alpha } e^{-3\tau }
),$$
\begin{equation} \label{EQ__26_}
Y\to \frac{1}{3} \sqrt{2\alpha } e^{-3\tau } ,\; H\to \frac{(2\alpha
)^{\frac{1}{4} } }{\sqrt{3\kappa } } e^{-\frac{3\tau }{2} } .
\end{equation}
For this equation there are three partial solutions $Y=0$, $Y=1/3$,
$Y=1/9$. Solution $Y=0$, $Y=1/3$ -- stable, $Y=1/9$ -- unstable.

At $\tau \to +\infty $ for $Y$ from an $(-\infty ,\; 1/9)$ interval
solution aspire at big $\tau $ to $Y=0$, $a(t)=const$, for $Y$ from
an $(1/9,\; +\infty )$ interval solution aspire big $\tau $ to
$Y=1/3$, $a(t)\to \alpha e^{\frac{t}{\sqrt{3\kappa } } } $.

By considering an interval of positive values $\tau $ at $\tau \to
+\infty $ we can observe a gradual weakening of the spinor field
influence, and within the limit we receive the prevalence of the
summands for a scalar field and the Einstein summand.

Note that in work Ref.  ~\refcite{lit22} solutions, with additional
asymptotic restrictions, for a conformal factor, in operation
without the spinor field are received. In our case, however, the
scalar field behaves differently, which is caused by the influence
of the spinor field summands. As the existence of a spinor excludes
a number of asymptotic solutions if we do not impose additional
approximations (for example degree). From the diagram (fig.
\ref{ris6})we can see that the expansion accelerates and then we see
deceleration. In a point 1.732 on an axis $Oa(t)$ the curve breaks
because further the function takes complex values.

\section{Conclusion}

The behavior of the cosmological model with the spinor field and
non-minimally interacted scalar field has been considered. In the
work a number of solutions addressing the exponential behavior of a
large-scale factor for model with a scalar and spinor field have
been calculated. That partial solution are solutions of $dS$
(de-Sitter), or solutions $a(t)=const$. Asymptotic solutions are
generally more difficult to comprehend and partial asymptotic
solutions are the de Sitter's solutions or solutions $a(t)=const$.
If we impose additional conditions on the asymptotic equations
(\ref{EQ__20_}), (\ref{EQ__23_}) it is possible to receive also the
series solutions for a large-scale factor. The existence of the
spinor field has an impact on evolution in the initial stage and
further weakens its influence. Towards the latter stages of
evolution, a scalar field and $R/2$ term are very influential. Our
following research will include the equivalent model with
interaction of the scalar field and $f(R)$ gravitation.

{\bf Acknowledgements.}

The work was partially supported by grant for leading Russian scientific schools, project No
88.2014.2 and grant of Ministry of Education and Science of Russian Federation, project No 867.


\end{document}